# Ultra-broad Band Signal Amplification Using Supercontinuum Generation

## Li Yikuan, Jiang Chun*

State Key Laboratory of Advanced Optical communication Systems and Networks, Shanghai Jiao Tong University, Shanghai 20040, China
*Corresponding author: cjiang@sjtu.edu.cn

**In this paper, ultra-broad band optical signal amplification are analyzed and demonstrated by utilizing supercontinuum generation propagating over the photonic crystal fiber. The coupled nonlinear Schrödinger equation containing parametric process and stimulated Raman scattering effect was analyzed and solved to calculate the variation of signal gain with frequency. The dependence of the gain spectra on nonlinear and chromatic dispersion coefficients as well as the geometrical parameters of photonic crystal fiber are discussed. The results show that when using a pump at wavelength 1450nm and peak power 1600W, the gain spectra covers from 1300 to 1700nm with the peak gain 100dB. This amplification mechanism have potential in full-wave optical fiber communication.**

http://dx.doi.org/10.1364/AO.99.099999

## 1. INTRODUCTION

A supercontinuum generation (SCG) is formed when a collection of nonlinear processes, such as Four-Wave Mixing (FWM), Cross-Phase Modulation (XPM), Raman scattering effect, act together upon a pump beam in order to cause severe spectral broadening of the original pump beam [1]. Owing to its distinct optical characteristics, SCG has been widely investigated both experimentally and theoretically in recent years. Currently, Photonic Crystal Fiber (PCF) is considered to be the exceptional waveguide for the realization of SCG. Surrounding the solid fused silica core with a regular array of microscopic air-filled holes, PCF is able to change its nonlinear and chromatic dispersion coefficients flexibly just by modifying the geometrical parameters [2]. A breakthrough research extended the spectral range of SCG from the violet to the infrared by propagating laser pulse with 100-fs duration and kilowatt-class peak power over PCF [3]. The researches concerning SCG keep progressing rapidly, while the applications of SCG are confined to optical fiber sensing technology [4] and Optical Coherence Tomography (OCT) [5]. Comparatively, the investigation of SCG application in ultra-broad band fiber transmission system is insufficient.

The traditional optical fiber communication system amplify the signal at particular intervals to compensate the loss. Generally, erbium-doped fiber amplifiers (EDFAs) and Raman fiber amplifiers (RFA) based on Stimulated Raman Scattering (SRS) effect are used [6]. However, EDFAs can just amplify the signal within the range 1530-1600nm, RFA needs large pump power with low energy efficiency, though [7,8].

In this paper, therefore, the possibility of amplifying ultra-broad band optical signal utilizing SCG over PCF is explored. By solving the coupled NLSE numerically, the signal gain spectra within the wavelength range 1250-1700nm at selected fiber length is calculated.

## 2. THEORETICAL ANALYSIS

### A. Signal Amplification Model

Previous researches about this model focus mostly on Raman gain caused by SRS and parametric gain caused by FWM. However, RFA can only amplify the signal whose wavelength is bigger than the pump's, while the gain of FOPA is not smooth enough. Consequently, Supercontinuum is resorted to make a contribution to the amplification model, which is not limited to these two defects.

SCG is formed by a collection of nonlinear processes that cannot be discussed separately. It is known that the SCG can be simulated by solving the generalized nonlinear Schrodinger equation (NLSE) of single pulse propagation [9] in optical fiber. It is expressed as follows:

$$\frac{\partial A}{\partial z} + \frac{\alpha}{2} A = \sum_{n \geq 2} \frac{i^{n+1}}{n!} \beta_n \frac{\partial^n A}{\partial T^n} + i\gamma (1 + \frac{i}{\omega_0} \frac{\partial}{\partial T}) \times$$
$$[A(z,T) \int_0^\infty R(T') | A(z, T-T') |^2 \, dT'] \qquad (1)$$

Where $A(z,T)$ is electric field envelope, which can derive the spectral intensity in both time and frequency domain; $A_{eff}$ -effective field envelope; $\omega_0$ -central pump frequency; $\alpha$ -fiber loss coefficient; c - speed of light in vacuum. Nonlinear coefficient $\gamma = n_2 \omega_0 / (c A_{eff})$, where $n_2$ is nonlinear refractive index, usually equals $3\times 10^{-20} m^2/W$ [10] in Silicon. $\beta_n$ is n-order dispersion coefficient. These two coefficients depend on the pump frequency $\omega_0$ as well as the PCF structures, which will be attentively discussed in the next section. The response function including both instantaneous electronic and delayed Raman contributions is $R(t)=(1-f_R)\delta(t)+f_R h_R(t)$. Where $f_R=0.18$, $h_R(t)$ represents delayed Raman contributions.

The numerical solution of Eq. (1), is carried out utilizing split-step Fourier scheme. In this method, the right-hand side of Eq.(1) is divided to linear and nonlinear parts, and integrated respectively. These two parts are combined together every fixed step length, expressed as Eq. (2), Where $\hat{D}$ and $\hat{N}$ are the linear and nonlinear parts:

$$A(z+h,T) \approx \exp(h\hat{N})\exp(h\hat{D})A(z,T) \quad (2)$$

The shorter the step is, the more accurate the integration is. Obviously, the nonlinear part is the convolution of $A(z,T)$ with $R(T')$. When calculating this part, the time domain is transformed to the frequency domain by using Fast Fourier Transform (FFT) [10].

What's more, when using SC source as a pump, extra nonlinear interactions such as IRS between the pump and signal laser have to be considered. The basic physical model when simulating the broadband signal amplification using SCG is that two beams of lasers propagate over the PCF simultaneously. Thus, the coupled NLSEs Eq. (3) contribute to the numerical model.

$$\frac{\partial A_{pump}}{\partial z}+\frac{\alpha}{2}A_{pump}=\sum_{n\geq 2}\frac{i^{n+1}}{n!}\beta_n^{pump}\frac{\partial^n A_{pump}}{\partial T^n}+i\gamma_{pump}(1+\frac{i}{\omega_0^{pump}}\frac{\partial}{\partial T})\times$$

$$[A_{pump}(z,T)\int_0^\infty R(T')(|A_{pump}(z,T-T')|^2+2|A_{signal}(z,T-T')|)dT']$$

**(3.a)**

$$\frac{\partial A_{signal}}{\partial z}+\frac{\alpha}{2}A_{signal}=\sum_{n\geq 2}\frac{i^{n+1}}{n!}\beta_n^{signal}\frac{\partial^n A_{signal}}{\partial T^n}+i\gamma_{signal}(1+\frac{i}{\omega_0^{signal}}\frac{\partial}{\partial T})\times$$

$$[A_{pump}(z,T)\int_0^\infty R(T')(|A_{pump}(z,T-T')|^2+2|A_{signal}(z,T-T')|)dT']$$

**(3.b)**

These coupled NLSEs ought to be used to numerically simulate multiple pumps SCG [11]. Multiple pumps can broaden the SCG spectral range remarkably. To describe interactions between the two beam (signal and pump) lasers, Eq. (3) compared with Eq. (1) takes cross-phase modulation term $2|A_i(z,T-T')|$ into consideration. The results in multiple pumps SCG simulation using Eq. (3) coincide with the experiments [12]. Hence, Eq. (3) has credibility in describing the model that two lasers propagate over the PCF simultaneously.

**B. Parameters of Photonic Crystal Fiber**

In the Eq. (1) and Eq. (3), nonlinear coefficient $\gamma$ and dispersion coefficients $\beta_n$ are both related to the pump frequency $\omega_0$ and the structure of PCF. Generally, these two sets of coefficients are obtained from experiment [13]. To simulate the broad-band signal amplification, all of these coefficients varying with wavelengths are required. Nevertheless, so numerous the workload is that achieving all these data by experiment is by no means easy.

PCFs can offer design flexibility by controlling the modal properties [14]. Usually, index-guiding PCFs consists of a central solid region, which mostly are silica, surrounded by multiple air holes in a regular triangular lattice.

In a PCF, $d$ -the air hole diameter, and $\Lambda$ -the hole pitch are both the geometrical parameters, while $n_2$ –non-linear refractive index, is the material parameter. A unique PCF which has a good performance in SCG can be obtained by modifying its geometrical parameters. Therefore, the empirical relations method, which has been validated highly consistent with the experimental data [14], are resorted to obtain $\gamma$ and $\beta_n$ varying with different wavelengths.

After approximate deduction, the relation can be written as [2, 14]:

$$\gamma = \frac{\omega_0}{\pi c^3}n_2 n_{co}(n_{co}-n_{eff}) \quad (4)$$

Where $n_{co}$ - refractive index of silicon dioxide, can achieve through Sellmeier equation [15]:

$$n_{co}^2(\omega_0)=1+\sum_{j=1}^m \frac{B_j \omega_j^2}{\omega_j^2-\omega_0^2} \quad (5)$$

Where $B_j$ and $\omega_j$ are empirical values. $n_{eff}$ is the effective index of the fundamental guided mode, which is achieved through Eq. (6):

$$n_{eff}^2 = n_{co}^2 + \frac{3(W^2-V^2)\lambda^2}{4\pi^2\Lambda^2} \quad (6)$$

$V,W$ are two dimensionless parameters, which are related to the empirical function of $(\frac{\lambda}{\Lambda},\frac{d}{\Lambda})$. The curves of $n_{eff}$, $n_{co}$ varying with wavelengths ranging from $0.5\mu m$ to $2\mu m$ with step of $1nm$ are obtained and shown in Fig. 1. That $n_{co}$ approximate to 1.45, compared with the experimental values, is convinced accurate. And $n_{eff}$ is also in the reasonable range.

Substituting Eq. (4) and (5) into Eq. (6), the variation of nonlinear coefficient $\gamma$ and dispersion parameter $D$ ,which is the first derivative of $\beta_1$ with respect to wavelength $\lambda$ and , included to describe the chromatic dispersion, can be achieved numerically and empirically.

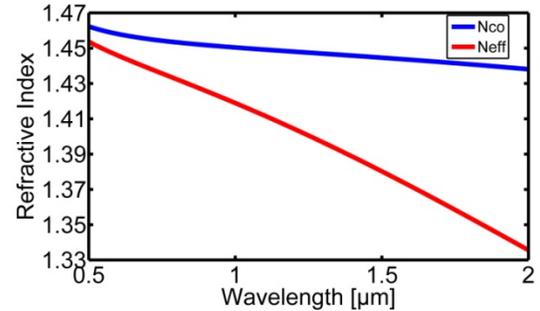

Fig. 1. Calculated results of refractive index of silicon dioxide (blue) and effective index of the fundamental guided mode (red) as a function of wavelengths from $0.5\mu m$ to $2\mu m$ with step of $1nm$ for $d=1.3\mu m$ and $\Lambda=1.7\mu m$.

In this paper, expansion of the dispersion coefficient $\beta$ into the Taylor series in frequency domain is carried out to the highest term with $n_{max}=5$. This is amply accurate to the simulation.

### C. Gain Spectra Calculation

Intensity of signal light which transmits in optical fiber communication system is quasicontinuous (light intensity for 0 bit is not zero), which is quite different from the origin usage of Eq. (3) that multiple pulsed pump propagating simultaneously. This condition causes several difficulties in the numerical simulation. A quasi-continuous signal laser can be seen as an infinite-duration pulse or a long-pulse when propagating over a short distance. According to the previous research, a method that using hundred-picosecond width time window to intercept a snapshot of quasicontinuous signal so as to simulate continuous pump stimulated SCG was proved to be effective [16,17].

The time-domain step length should conform to Nyquist Criterion that the minimum sampling frequency should be twice of the effective sinusoidal frequency at least. Big step may result in displaying spectrum fragmentarily. Smaller step length can obtain clearer numerical results, while leading to bigger computational complexity. Hence, based on the above consideration, 1fs is adopted as the step length, and a 262-ps width time window is used.

Additionally, when setting the initial injected laser, random noise is loaded as the phase diffusion model mentioned et al. [18]. This model consists of a Gaussian shape with randomly spectral phase. Although the physical justifications are being explored, this approach is used since its excellent agreement between the experimental and numerical results. The noise expression in frequency domain is shown as Eq. (7) [19]:

$$A_{noise}(\nu_m) = \sqrt{E_{photon}} \times \exp[i\phi(\nu_m)] \quad (7)$$

Where $A_{noise}(\nu_m)$ -the power of injected photon, $\phi(\nu_m)$ -the random phase of photon.

## 3. Result and Discussion

The parameters of PCF are designed to be the hole diameter $d=0.575\mu m$, hole pitch $\Lambda=2.5\mu m$. The fiber with these couple of parameters can generate a series of ultra-high nonlinear coefficients around $\gamma=0.01W^{-1}m^{-1}$ that can result in a broad-band SCG; and an ideal ZDW (Zero-dispersion Wavelength) value around $\lambda_{ZDW}=1405nm$, which is the center wavelength of communication band shown in Fig. 2.

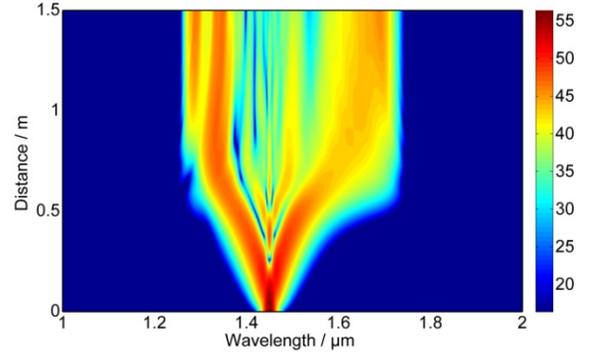

Fig. 2. Results of nonlinear coefficient(blue) and dispersion coefficient(green) as a function of wavelength from $0.5\mu m$ to $2\mu m$ with step of $1nm$ for $d=0.575\mu m$ and $\Lambda=2.5\mu m$. Zero dispersion wavelength is close to 1405nm.

Next, SCG is generated numerically. Considering a $1.5m$ long PCF, whose structure is designed as above. The initial injected pump source has a hyperbolic secant field profile $A(0,T)=\sqrt{P_0}\operatorname{sech}(T/T_0)$ with ultra-high peak power $p_0=1.6kW$, $FWHM=176ps$ [20].

With the parameters above, the Eq. (3) is solved numerically by using split-step Fourier scheme. Fig. 3 presents the result of propagation. This Pseudo-color map illustrates the logarithmic scale intensity of wavelengths ranging from $1\mu m$ to $2\mu m$. Due to the high nonlinearity of PCF as well as ultra-high pump power, the spectral range broaden promptly to $0.5\mu m$ from a pulse over the very first $0.5m$ transmission. Then the spectra evolution is stabilized.

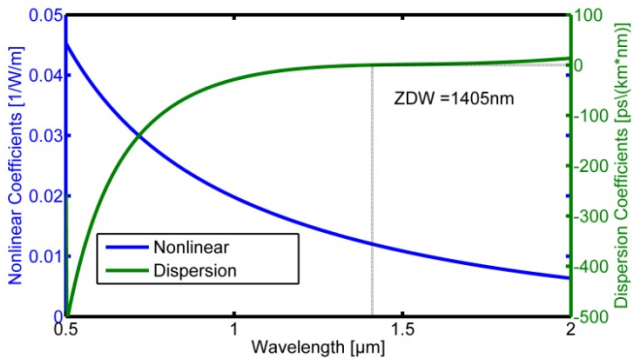

Fig. 3. Results from numerical simulations showing spectral evolution at selected propagation distances. The sech2 input pulse at 1450 nm has 1.6kW peak power and 176ps FWHM. Fiber parameters are given in the text of this section. Unless other stated, the 40dB dynamic range density scale down applies to all density plots.

Considering Fig.3, the internal FWM and XPM is remarkable so that such a broad continuum is achieved. When injecting another beam, the interaction between these two beams will enhance the nonlinear processes.

Next, signal laser is injected into the fiber. The signal laser involved is an continuous-source with power $p_{signal}=1mW$ loading random phase as mentioned above. To discuss the dependence of the signal intensity gain on the signal wavelength variation, the wavelengths are set embracing the band from 1225 to 1725nm in steps of 50nm. The transmission loss is assumed to be $1\times10^{-4}dB/m$. Eq. (8) is involved to evaluate the gain at certain signal wavelength.

$$G_{gain} = 10\lg\frac{P_{out}}{p_{in}} \quad (8)$$

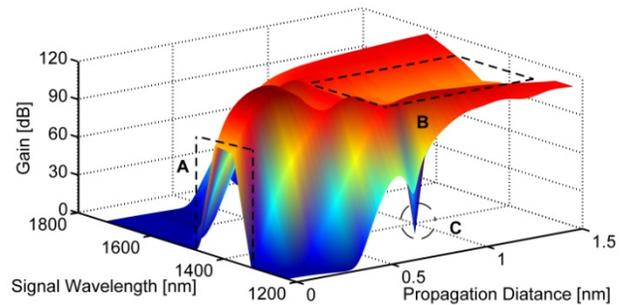

Fig. 4. Results of signal gain evolution in log-scale at selected propagation distances. The continuous signals have 1mW power, covering 1200 to 1800nm with step size 1nm. Fiber parameters are given as above. Three special gain zones are tagged in the figure. A: Initial gain gate; B: Smooth gain band; C: Abnormal gain.

From Fig. 4, it is perceived that the signal within the selected band are amplified to 100dB averagely. The signal wavelengths within 1300-1500nm tagged as Zone (B) achieve smoother gain and better result compared with other bands. That SCG spectrum expands from its central frequency results in this smoother segment. Another factor is that the SRS process is more notable if the frequency gap between pump and signal is smaller.

Fig. 5 shows three signal at different bands. The signals at $1375nm$ and $1525nm$ (Zone B) reveal a good amplification. The $1525nm$ signal occurs to a distance of monotonically increasing before stabilization. However, the $1375nm$ signal has an ultra-big gain at the very beginning, being different from other bands. Zone (A), an initial gate region tagged in Fig. 4&5, reflects this phenomenon. This is because the wavelength gap between signal and pump is so small that the injected pump power covering the signal contribute to this abnormal gain.

The gain will converge to $100dB$ within the communication-band after propagating half a meter. On the contrary, the intensity of $1\mu m$ signal is keep decreasing, for SC don't expand to this band and signal experiences background loss.

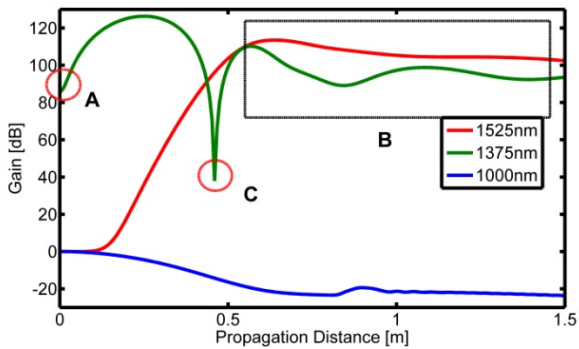

Fig. 5. Results of signal gain as a function of propagation distance at different wavelength of 1525(red), 1375(green) and 1000nm(blued). Three special gain zone are also tagged. A: Initial gain gate; B: Smooth gain band; C: Abnormal gain.

Zone(C) is tagged in both Fig.4 and Fig.5. Some signals, where is near the pump wavelength, suffer a short distance of abnormal or negative gain when propagating. Although the signal will lost gradually, the attenuation could not be so intense. The reason is that not a single nonlinear process is responsible for the amplification. Two predominant effects, FWM and XPM, have uncertainty. Generally, they will bring about notable gain. But, abnormal gain will randomly occur at certain distance and wavelength where the nonlinear processes are absent. As the figure reflects, Zone(C) emerges. With the evolution being, this possibility will decreased significantly, and a smooth gain emerges.

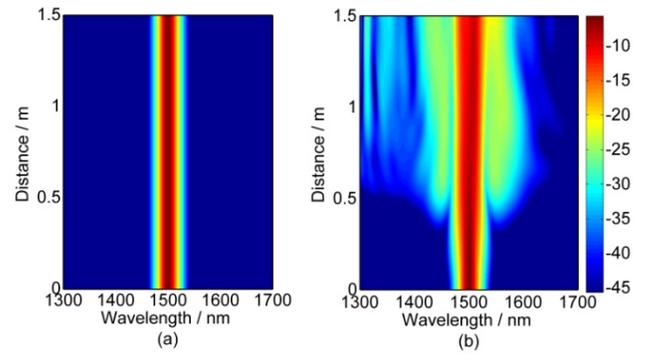

Fig. 6. Results from numerical simulations showing spectral evolution at selected propagation distances. Fig(a). 1500nm-1mW signal transmits over the fiber individually. Fig(b) the same parameters signal propagates only ignoring the process of pump laser broadening. The pump source of Fig(b) shares the same parameters with above. the 40dB dynamic range density scale down is involved.

The scheme of amplification is not the direct addition of SCG and signal transmission. The amplified signal consists of three parts. One part is the signal laser itself. Another part is that the nonlinear processes, such as FWM (Four-Wave Mixing), XPM (Cross-Phase Modulation), act together upon the pump laser, the pump source expands to the signal band. This term shares the same principle with SCG. Moreover, the third part of amplification emerges like the FRA. Pump resource is affected by SRS, several energy of pump laser was stimulated to the signal wavelength. The cross modulation term $2|A_i(z,T-T')|$ in Eq. (3) describes the numerical process of this factor. The comparison between the signal intensity whether pump source is involved can also demonstrate the interaction between SC and signal transmission. The signal wavelength is set at $\lambda_{signal}=1500nm$.

Fig. 6(b), deriving from $A_{signal}$ in Eq. (3), reveals the signal intensity propagation only ignoring the process of pump laser broadening. While Fig. 6(a) shows the condition that only signal laser transmits over the fiber without the stimulation action of pump resource. Compared these two pseudo-color maps, it can be seen that the pumped signal changes a lot. Hence, interaction is an important factor contribute to the gain.

## 4. Conclusion

In this paper, ultra-broad band signal amplification was simulated by adopting entirely numerical approaches. A unique PCF structure, whose ZDW was near the central wavelength of communication band, was designed and solved numerically using empirical method. It is concluded that SCG pump source can remarkably amplify the optical signal covering communication band.

This paper put forward a reasonable theoretical expression of the amplification scheme. This amplification mechanism would be applied to full-wave optical fiber communication, and the development in new type optical fiber amplifier.

There are still many works to be continued, such as the selection of pump wavelength and the noise figure evaluation. We look forward to obtaining more smooth gain spectra covering full-wave communication window with the aid of SC generation in our future researches.